\documentclass[multphys,vecphys]{svmult}

\usepackage{makeidx}         % allows index generation
\usepackage{graphicx}        % standard LaTeX graphics tool
\usepackage{multicol}        % used for the two-column index
\usepackage[bottom]{footmisc}% places footnotes at page bottom
\makeindex             % used for the subject index
\begin{document}

\title*{Tracing Ghost Cavities with Low Frequency Radio Observations}
% Use \titlerunning{Short Title} for an abbreviated version of
% your contribution title if the original one is too long
\author{T.~Clarke\inst{1,}\inst{2}\and E.~Blanton\inst{3}\and C.~Sarazin\inst{4}\and N.~Kassim\inst{1}\and L.~Anderson\inst{3}\and H.~Schmitt\inst{1,2}\and Gopal-Krishna\inst{5}\and D. Neumann\inst{6}}
%\and
%Name of Author\inst{2}}
\authorrunning{Clarke et al.}
% Use \authorrunning{Short Title} for an abbreviated version of
% your contribution title if the original one is too long
\institute{Naval Research Laboratory, 4555 Overlook Ave SW, Washington, DC USA
\and Interferometrics, Inc., 13454 Sunrise Valley Drive, Herndon, VA USA
\texttt{tracy.clarke@nrl.navy.mil}
\and Boston University, 725 Commonwealth Ave., Boston, MA USA
\and University of Virginia, 530 McCormick Rd., Charlottesville, VA USA
\and NCRA-TIFR, Pune University Campus, Pune India
\and CEA/Saclay, L'Orme des Merisiers, Gif-sur-Yvette, France}
%\and Name and Address of your Institute \texttt{name@email.address}}
%
% Use the package "url.sty" to avoid
% problems with special characters
% used in your e-mail or web address
%
\maketitle

\begin{abstract}
We present X-ray and multi-frequency radio observations of the central
radio sources in several X-ray cavity systems. We show that targeted
radio observations are key to determining if the lobes are being
actively fed by the central AGN. Low frequency observations provide a
unique way to study both the lifecycle of the central radio source as
well as its energy input into the ICM over several outburst episodes.
\end{abstract}

\section{Introduction}
\label{sect:intro}

The radiative cooling time in the central regions of many dense
clusters is less than the age of the cluster. Without any outside
disturbance this gas should continue to cool to very low temperatures
creating a ``cooling flow'' in the cluster center \cite{fabian94}. The
search for cool gas through X-ray spectroscopic observations has
revealed the surprising fact that the temperature drop in cluster
cores seems to halt at temperatures around one-third of the maximum
cluster temperature \cite{peterson06}. An obvious candidate for energy
input to offset significant cooling is the cluster-center AGN, which is
typically radio-loud in dense cooling core systems.

X-ray images of several cooling core clusters reveal the presence of
significant interaction between the central radio source and the
intracluster medium (ICM), thus supporting the idea that the AGN may
be a source of energy input to the surrounding gas. The ICM shows
evidence of cavities surrounded by (at least partial) rims in several
systems. Associated radio observations at 1.4 GHz reveal that many of
the cavities are filled with radio plasma from the lobes of the
central AGN. In some systems, however, the X-ray cavities are not
associated with emission at this frequency. These cavities are
generally referred to as ``ghost cavities'' and are thought to be the
result of buoyantly rising lobes from past radio outbursts. One such
system, Perseus \index{Perseus}\cite{fabian04}, shows low frequency radio spurs
toward the ghost cavities which suggests that at least some ghost
cavities are filled with old radio plasma. In order to understand the
details of the AGN energy input into the ICM is it necessary to
combine the X-ray observations with targeted radio observations over a
range of different frequencies. Here we present a few examples of new
results from multi-frequency radio observations of cavity
systems.

\section{Abell 4059 \index{A4059}}
\label{sect:a4059}

\begin{figure}[t]
\centering
\includegraphics[height=3.34cm]{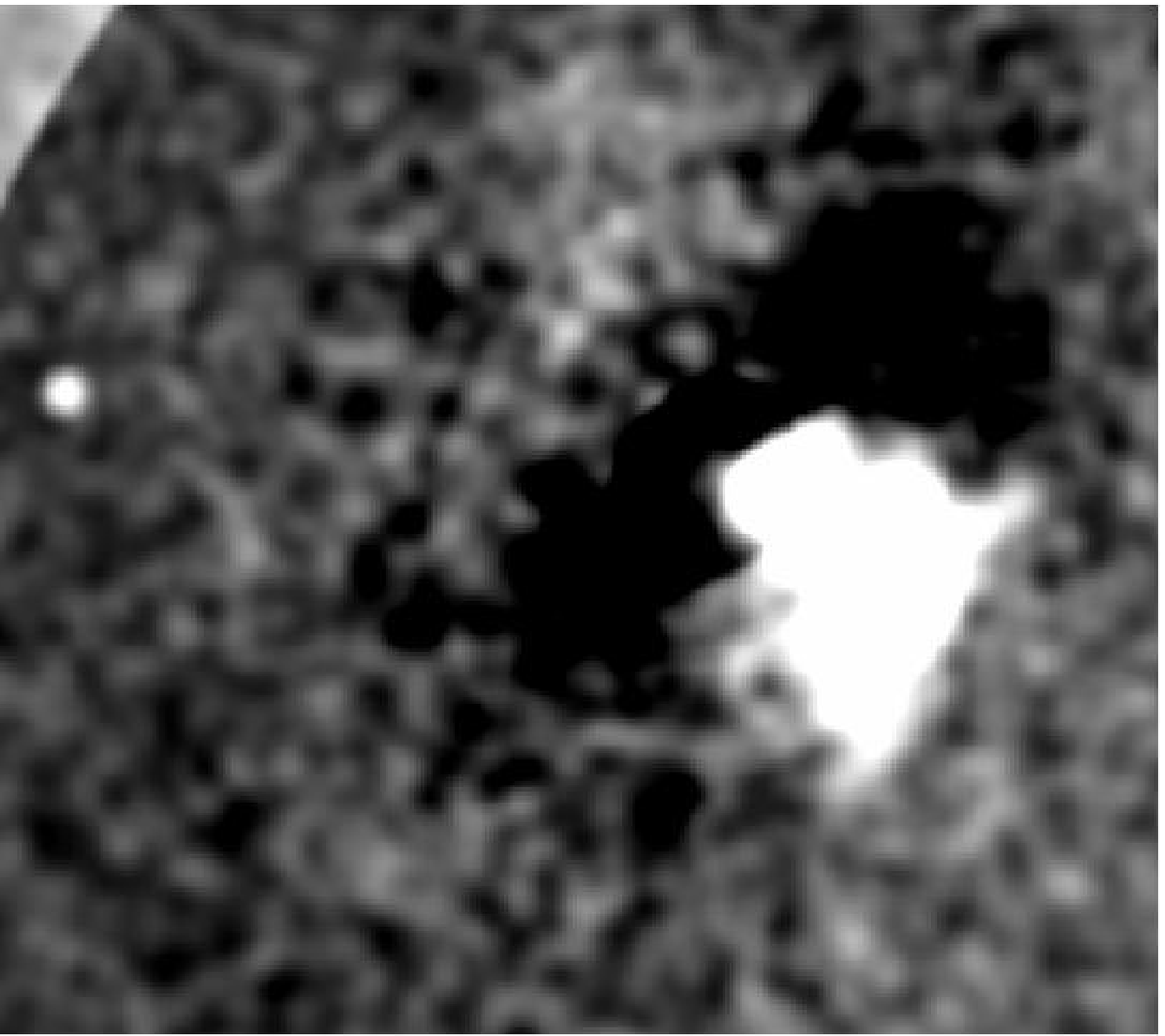}
\includegraphics[height=3.34cm]{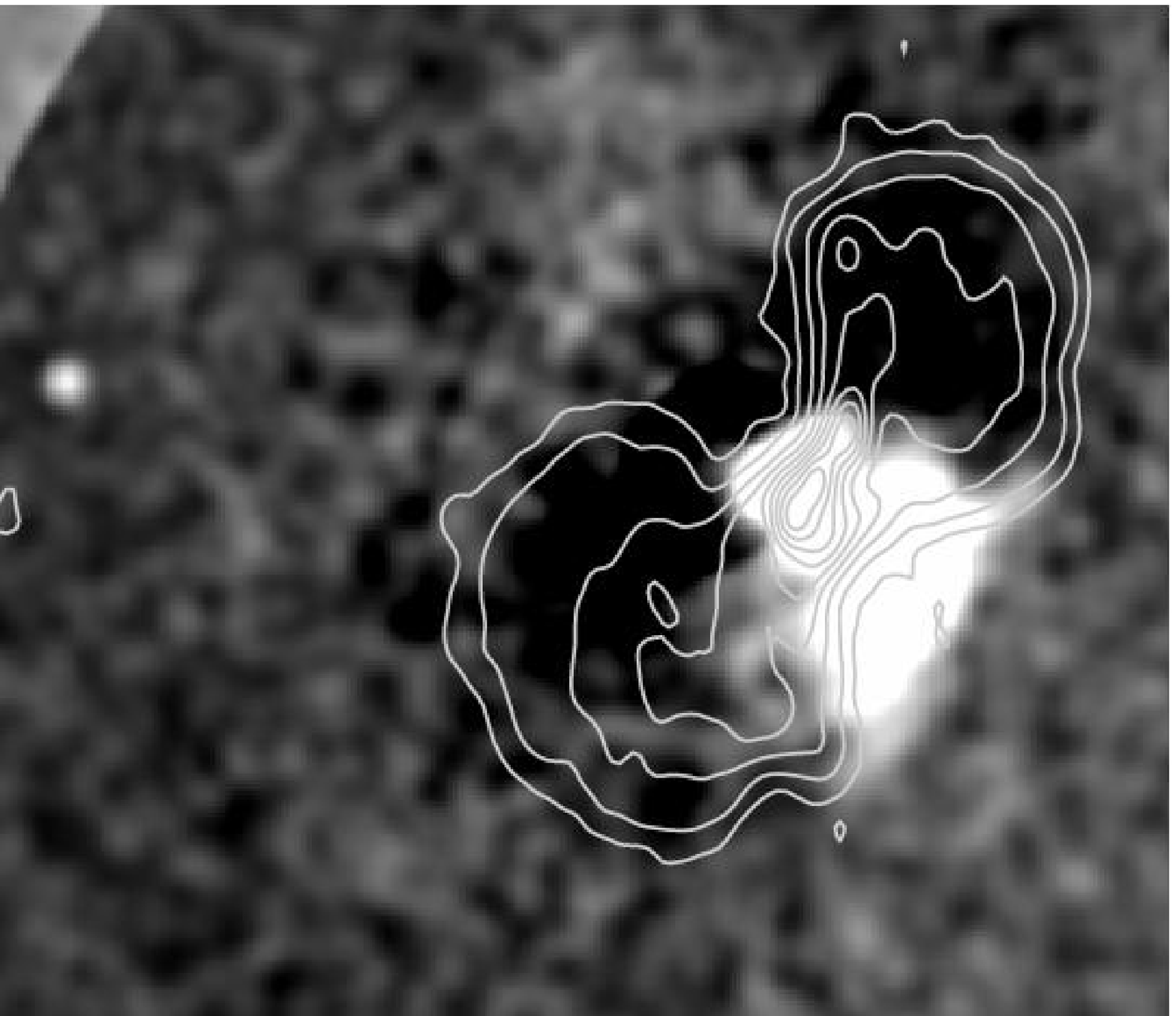}
\includegraphics[height=3.34cm]{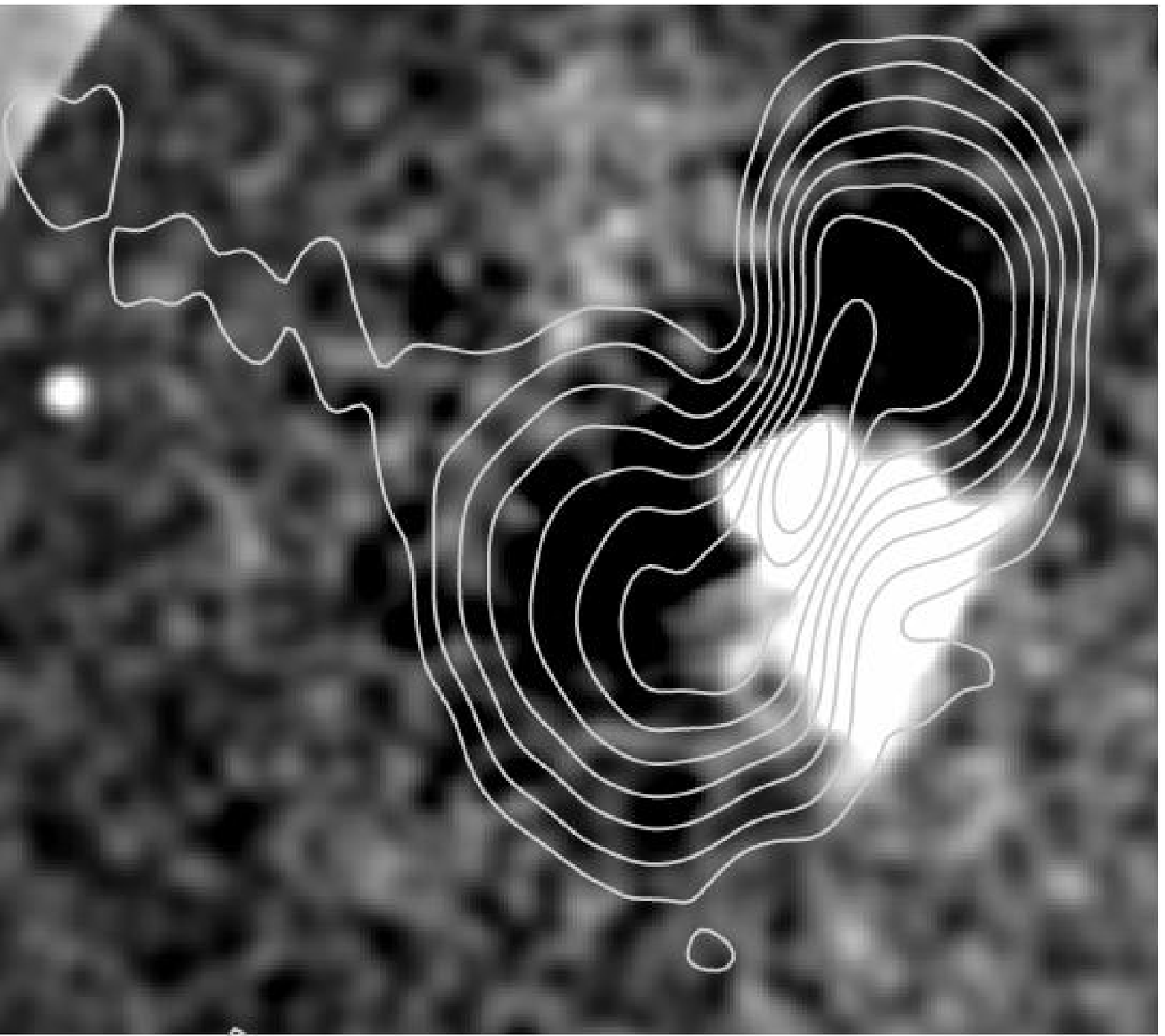}
\caption{{\bf Left} Unsharp masked $Chandra$ image of the central
$\sim 120 \times 150$ kpc region of Abell 4059 showing the X-ray bar
across the cluster center and the two cavities located to the N and SE
of the cluster core. {\bf Middle} VLA 1.4 GHz contours overlaid on the
$Chandra$ image. The radio lobes bend to completely fills both
cavities. {\bf Right} VLA 330 MHz radio contours show somewhat larger
lobes and a possible trail of emission running to the NE of the
eastern lobe.}
\label{fig:a4059}    
\end{figure}

X-ray cavities in Abell 4059 were first detected with the High
Resolution Imager on the ROSAT satellite \cite{huang98}. Subsequent
observations with $Chandra$ revealed the presence of a central X-ray
bar and showed that the two cavities (Figure~\ref{fig:a4059} left)
were not symmetric about the cluster center \cite{heinz,choi}. Radio
observations at 4.8 and 1.4 GHz showed that the emission associated
with the central AGN was not aligned with the two cavities and did not
extend as far as either cavity \cite{choi}. In this context the
cavities in Abell 4059 were thought to be buoyant ghost cavities from
a previous outburst.

New VLA observations of Abell 4059 at frequencies of 1400 and 330 MHz
show that the X-ray cavities are filled with radio plasma
(Figure~\ref{fig:a4059} middle \& right). The radio emission to the
north extends beyond that seen by \cite{choi} and bends to the west to
fill the cavity. Similarly the southern radio lobe bends to the east
to fill the other X-ray cavity. In fact, radio emission within the
cavities is seen at frequencies as high as 4.8 GHz, suggesting that
the cavities are still being actively fed by the central AGN.

\section{Abell 2597 \index{A2597}}
\label{sect:a2597}

\begin{figure}[t]
\centering
\includegraphics[height=3.9cm]{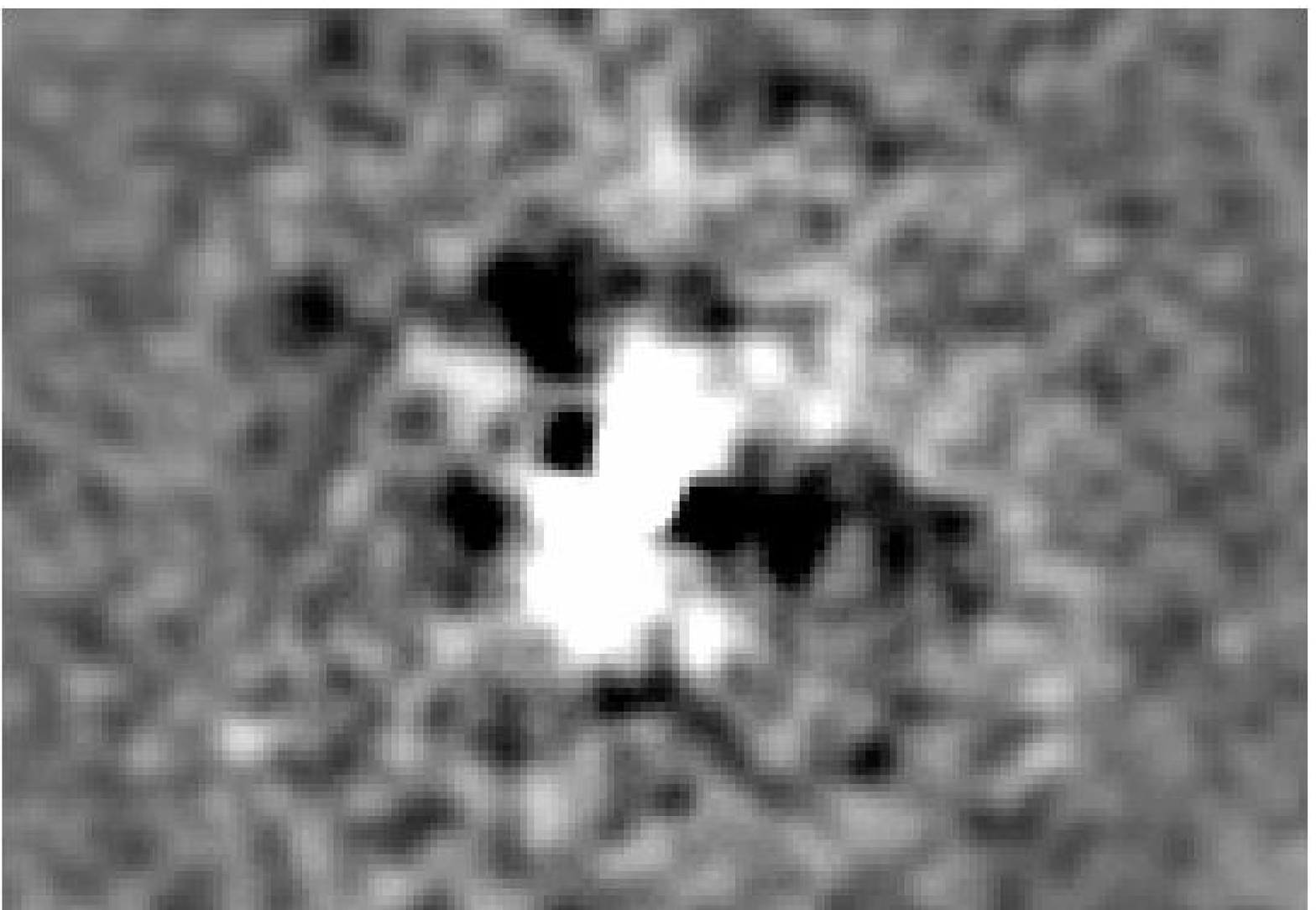}
\includegraphics[height=3.9cm]{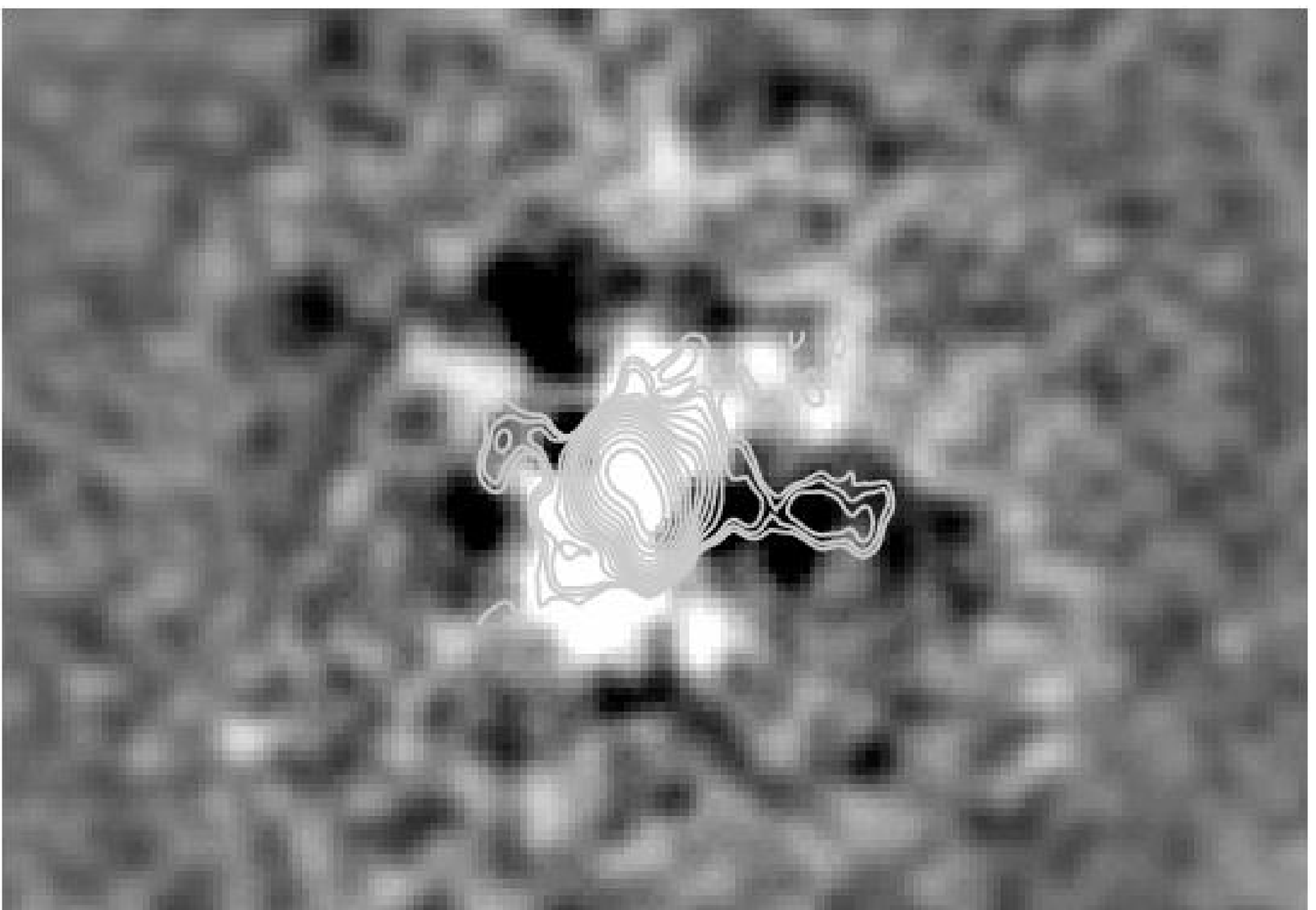}
\caption{{\bf Left:} Unsharp masked $Chandra$ image of the central
$\sim 85 \times 125$ kpc region of Abell 2597. This deep (112 ks)
image confirms the X-ray tunnel and cavities to the NE and shows
evidence of a bright rim surrounding the tunnel. {\bf Right} VLA 1.4
GHz radio contours overlaid on the unsharp masked $Chandra$ image.}
\label{fig:a2597}    
\end{figure}

Abell 2597 is well known to host a compact C-shaped radio source in
the cluster core \cite{sarazin95}. X-ray observations reveal the
presence of ghost cavities \cite{mcnamara01} as well as a tunnel in
the ICM gas connecting the core to the western ghost cavity
\cite{clarke05}. Evidence for cool gas is seen in HST FUV observations
which show diffuse emission as well as filaments and knots
\cite{O'Dea}, while further evidence from cool gas in the cluster core
comes from FUSE OVI observations \cite{oegerle}.

Radio observations of the cluster core at frequencies of 4.8 GHz and
below reveal emission extended beyond the compact C-shaped source. Low
frequency (330 MHz) observations show synchrotron emission filling
the tunnel \cite{clarke05} while at 1.4 GHz (Figure~\ref{fig:a2597}
right) the emission appears to be clumpy and only fills a portion of
the tunnel. The 1.4 GHz radio contours also trace emission extended
to the NW, as well as an arc of emission to the NE which is associated
with the inner NE cavity as well as a bright Ly$\alpha$ filament seen
in the HST images.

\section{Abell 262 \index{A262}}
\label{sect:a262}

Radio observations of the core of Abell 262 show that it is host to a
weak double lobed source \cite{parma86} which displays an S-shaped
morphology. This radio source appears to be interacting with the
surrounding thermal gas as seen from $Chandra$ observations of the
system which revealed an X-ray cavity associated with the eastern
radio lobe \cite{blanton04}. An analysis of the energy required to
create the X-ray cavity compared to the cooling luminosity in the
system suggested that the current outburst in Abell 262 is too weak to
provide sufficient energy to offset cooling \cite{blanton04}.

\begin{figure}[t]
\centering
\includegraphics[height=3.9cm]{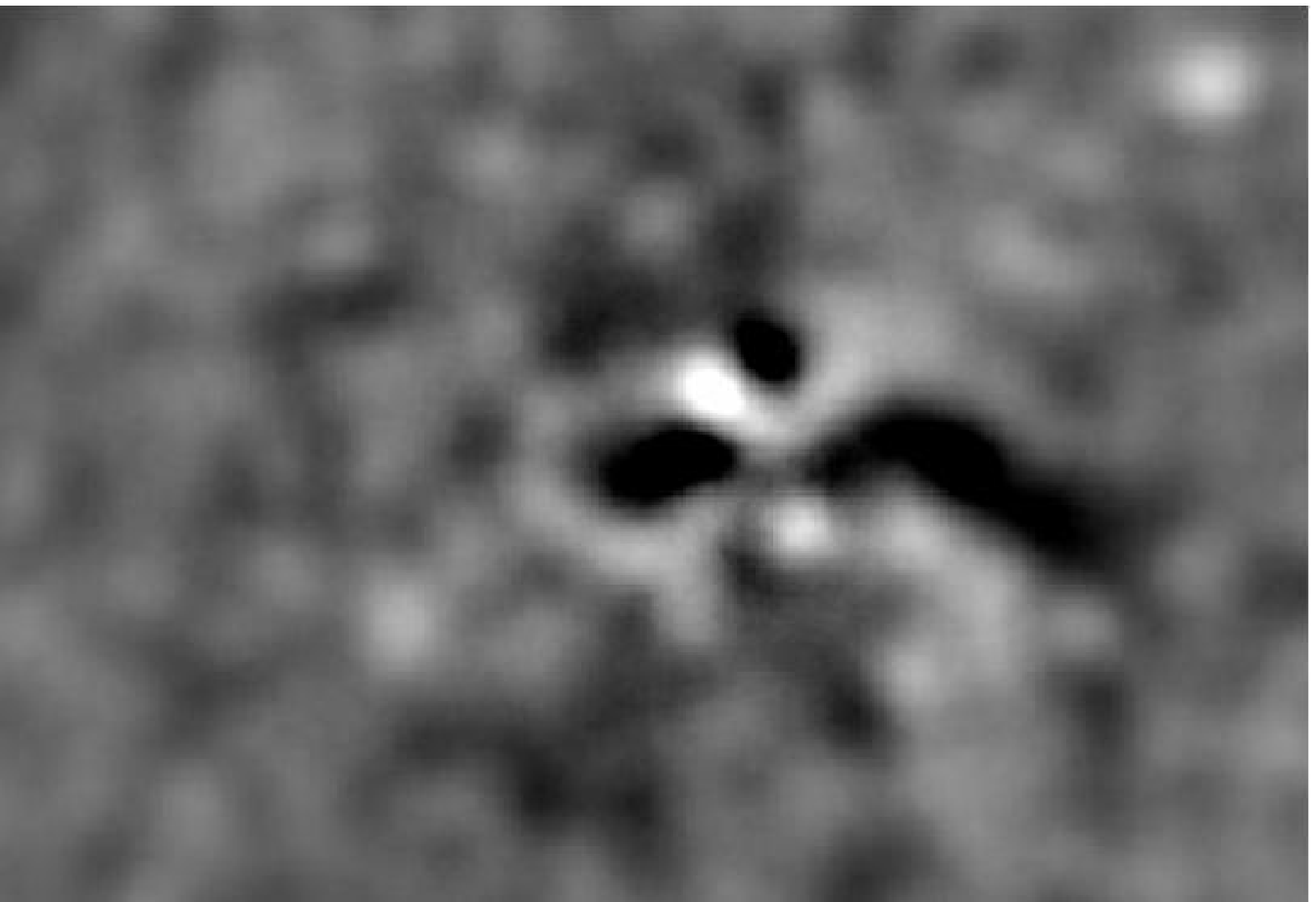}
\includegraphics[height=3.9cm]{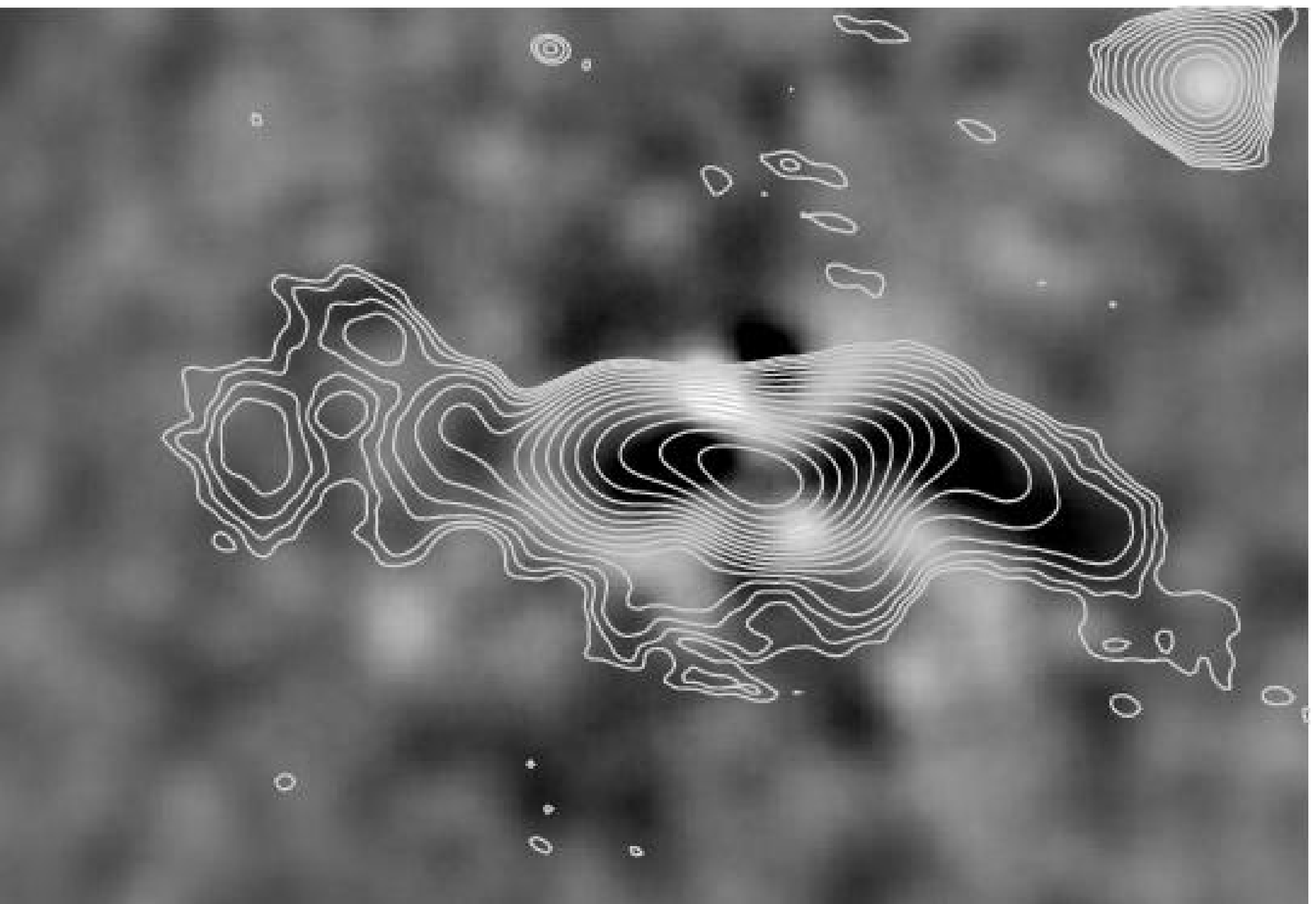}
\caption{{\bf Left:} Unsharp masked $Chandra$ image of the central
$\sim 50 \times 80$ kpc region of Abell 262 showing the eastern cavity
as well as the tunnel to the west of the cluster core. {\bf Right:}
GMRT 610 MHz contours of the radio emission from the central AGN in
Abell 262 overlaid on the $Chandra$ residual image.}
\label{fig:a262}    
\end{figure}

Our unsharp masked analysis of the $Chandra$ data of Abell 262
\cite{anderson} reveals that the eastern cavity is surrounded by a
complete X-ray rim. In addition, the cluster is also host to an X-ray
tunnel running westward from the AGN (Figure~\ref{fig:a262}
left). Multi-frequency radio observations at frequencies below 1.4 GHz
reveal that the central radio source is more than three times larger
than previous observations showed \cite{clarke}. The 610 MHz emission
(Figure~\ref{fig:a262} right) fills the western tunnel, eastern X-ray
cavity, and reveals three distinct radio features further eastward of
the previously detected X-ray cavity. In fact the radio feature
closest to the eastern X-ray cavity falls on top of a low significance
X-ray deficit seen in the original $Chandra$ images
\cite{blanton04}. The other more eastern radio features are not
detected as X-ray deficits but this may simply be a result of having
insufficiently deep X-ray images to detect the depression. If the
observed radio features all correspond to separate AGN outbursts
(repetition timescale $\sim 3\times 10^7$ yr) that have created X-ray
cavities, then we find that the total energy input from those
outbursts is within a factor of two of the X-ray cooling luminosity.

\section{NGC 507 \index{NGC507}}
\label{sect:n507}

\begin{figure}[t]
\centering
\includegraphics[height=7cm]{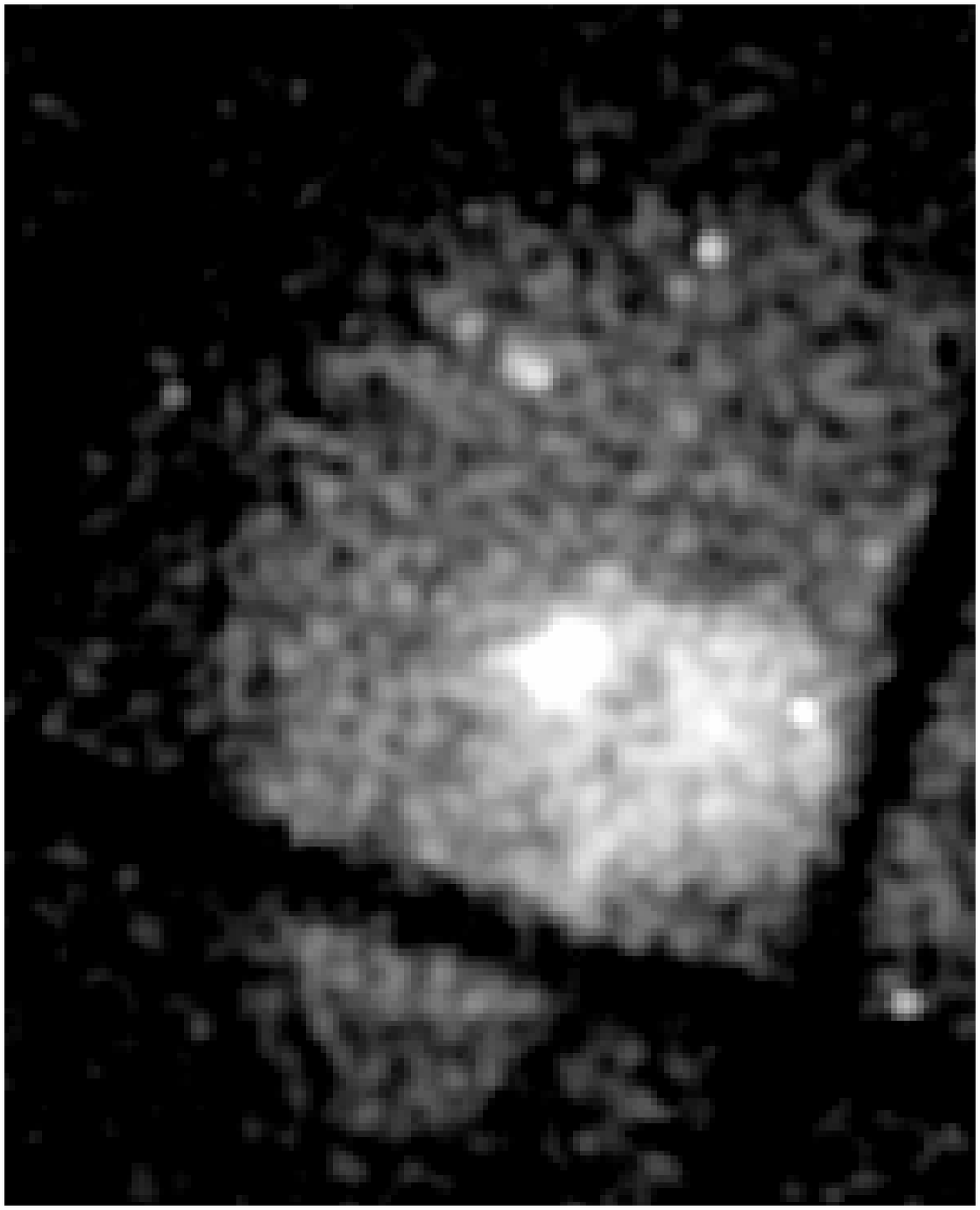}
\includegraphics[height=7cm]{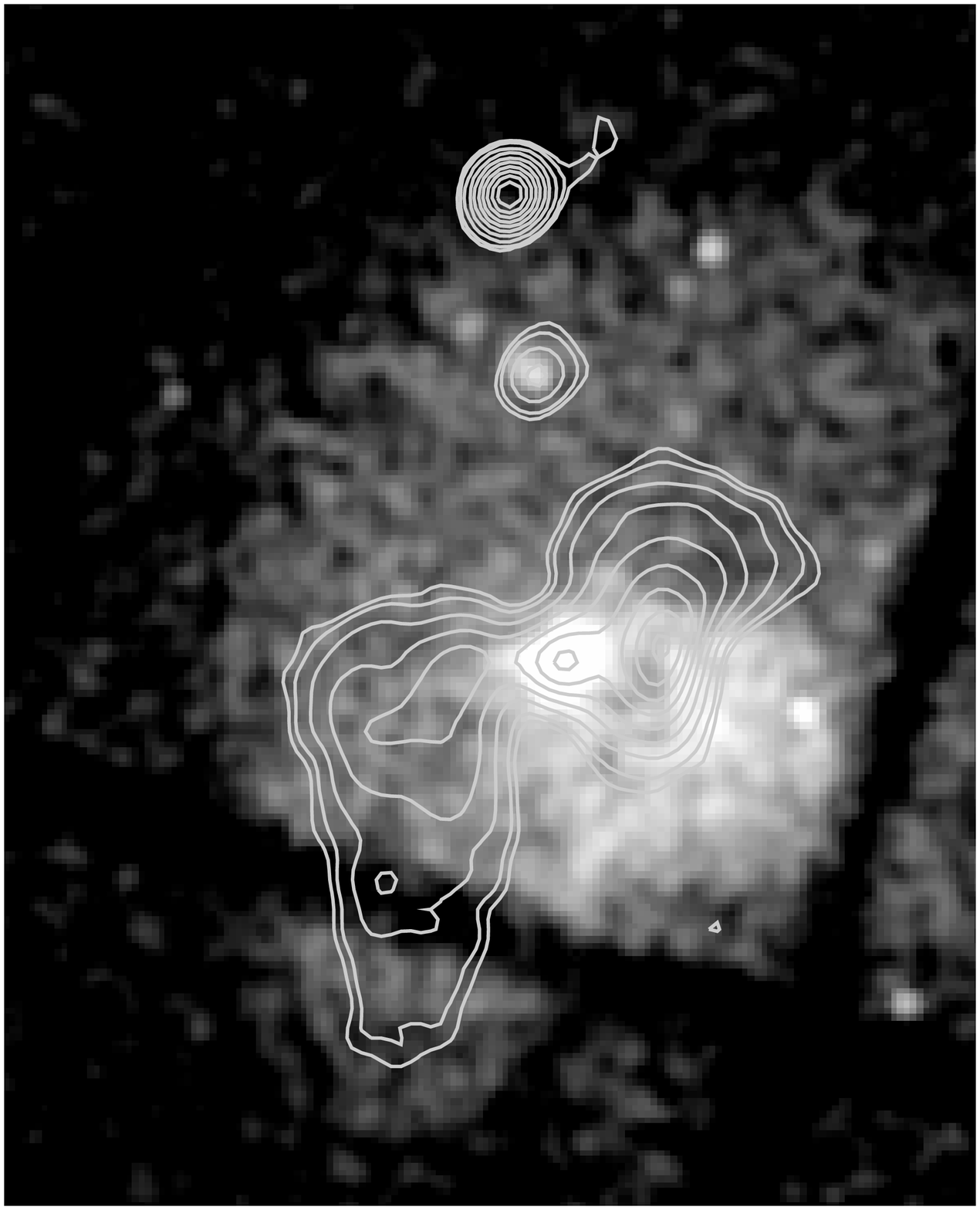}
\caption{{\bf Left:} Gaussian smoothed $Chandra$ X-ray image of the
central $\sim 115 \times 90$ kpc region of NGC 507. The thermal gas is
separated into two clumps in the core and a sharp-edged extension
running from the NE to the SE. {\bf Right:} VLA 1.4 GHz radio contours
show that radio emission traces morphology of the X-ray surface
brightness edge to the SE.}
\label{fig:n507}    
\end{figure}

The galaxy group NGC 507 displays significant evidence of disturbance
to the central X-ray emission \cite{kraft04}. The core shows two
central X-ray clumps surrounded by an extended diffuse emission which
displays a sharp surface brightness edge running from the northeast to
the southeast (Figure~\ref{fig:n507} left). The sharp edge may be
related to a metallicity gradient in the gas where the brighter
emission is associated with cooler, higher abundance material
\cite{kraft04}. A suggested origin for this material is gas which has
been displaced from the central regions by the expanding radio lobe
\cite{kraft04}. In the right panel of Figure~\ref{fig:n507} we show
new high resolution VLA 1.4 GHz contours overlaid on the X-ray surface
brightness image. The eastern radio lobe is seen to take a sharp bend
to the south to trace the inner edge of the X-ray discontinuity, while
the western radio lobe seems to be strongly interacting with the
central ICM, creating a possible depression between the two central
X-ray peaks.

\section{Discussion}
\label{sect:disc}

There is no question that the observations obtained from $Chandra$ and
XMM have significantly advanced our understanding of the interactions
between the central AGN and the thermal gas in galaxy groups and
clusters. It is clear that over a wide range of system masses the
central AGN has the ability to disrupt the thermal gas in the dense
cores of these systems. A detailed understanding of these central
interactions requires the addition of targeted multi-frequency radio
data that is selected to probe both the relevant spatial scales for
the system in question as well as a wide frequency range (including
low frequencies) to track the radio outburst history of the central
AGN.

In this paper we presented a few cases of sources where new radio
observations have changed the view of the system under study. In the
case of Abell 4059 it appears that the X-ray cavities are likely still
undergoing active injection from the central AGN and thus are not
buoyantly rising detached lobes. The radio observations of the central
source in Abell 2597 reveal extended structures along several
different position angles which are suggestive of multiple outbursts
along possibly different initial directions. 

In the case of Abell 262 the new radio observations reveal multiple
structures along roughly the same position angle. These distinct
structures may be signatures of different outburst episodes. Adding up
the total energy input into the ICM from all radio outbursts seen in
Abell 262 suggests that the AGN is more powerful than previously
thought and is within a factor of two of being powerful enough to
offset cooling over several outburst episodes.

The presence of the extended radio emission toward the eastern region
of Abell 262 suggests that the low frequency radio flux is originating
from detached buoyant lobes from past AGN outbursts. Using the
observed projected offset of the radio structures from the cluster
cores in Abell 2597 and Abell 262 together with the buoyant velocity
in the systems we estimate a repetition timescale for the central AGN
of a few $\times 10^7$ yr. This is significantly shorter than the
typical assumption of $10^8$ yr. Although there is some evidence of an
X-ray deficit associated with one of the eastern emission regions in
Abell 262 we note that it is relatively difficult to detect X-ray
cavities once they are located beyond the densest parts of the cluster
cores \cite{ensslin02}. In these cases, low frequency radio
observations provide an important tool for tracing the total energy
input into the system to determine if it is sufficient to offset the
radiative cooling.

Finally, in the case of NGC 507 we note that sensitive, high
resolution radio images allow us to trace the interactions of the
western radio lobe with the central X-ray structure. These
observations also show the clear correspondence between the eastern
radio lobe and the southern portion of the sharp surface-brightness
discontinuity in the system.

T.~E.~C. acknowledges support from NASA through $Chandra$ award
GO6-7115B. Basic research in radio astronomy at the Naval Research
Laboratory is supported by 6.1 base funding.
%
%
% BibTeX users please use
%\bibliographystyle{}
%\bibliography{paper}
%
% Non-BibTeX users please follow the syntax
% the syntax of "referenc.tex" for your own citations
%\input{osullivan_ref}

%%%%%%%%%%%%%%%%%%%%%%%%%%%%%%%%%%%%%%%%%%%%%%%%%%%%%%%%%%%%%%%%%%%%%%  }

%%%%%%%%%%%%%%%%%%%%%%%%%%%%%%%%%%%%%%%%%%%%%%%%%%%%%%%%%%%%%%%%%%%%%%

\printindex

\end{document}